\documentclass[bst/sn-mathphys,iicol]{sn-jnl}% Math and Physical Sciences Reference Style
\usepackage{lineno}
\usepackage{bm}
\usepackage{physics}
\usepackage{amsmath}
\usepackage{soul}

\modulolinenumbers[5]
\graphicspath{{./Latex/figures/}}

\jyear{2022}%

\theoremstyle{thmstyleone}%
\theoremstyle{thmstyletwo}%

\theoremstyle{thmstylethree}%

\newcommand{\pfrac}[2]{\frac{\partial #1}{\partial #2}}

\begin{document}

\title[Article Title]{Direct van der Waals simulation (DVS) of phase-transforming fluids}


\author{\fnm{Tianyi} \sur{Hu}}\email{hu450@purdue.edu}

\author{\fnm{Hao} \sur{Wang}}\email{wang5336@purdue.edu}

\author*{\fnm{Hector} \sur{Gomez$^*$}}\email{hectorgomez@purdue.edu}

\affil{\orgdiv{School of Mechanical Engineering}, \orgname{Purdue University}, \orgaddress{\street{585 Purdue Mall}, \city{West Lafayette}, \postcode{49706}, \state{Indiana}, \country{USA}}}

%\affil[2]{\orgdiv{Weldon School of Biomedical Engineering}, \orgname{Purdue University}, \orgaddress{\street{585 Purdue Mall}, \city{West Lafayette}, \postcode{49706}, \state{Indiana}, \country{USA}}}


\abstract{We present the method of Direct van der Waals simulation (DVS) to study computationally flows with liquid-vapor phase transformations. Our approach is based on a novel discretization of the Navier-Stokes-Korteweg equations, that couple flow dynamics with van der Waals' non-equilibrium thermodynamic theory of phase transformations, and opens an opportunity for first-principles simulation of a wide range of boiling and cavitating flows. The proposed algorithm enables unprecedented simulations of the Navier-Stokes-Korteweg equations involving cavitating flows at strongly under-critical conditions and $\mathcal{O}(10^5)$ Reynolds number. The proposed technique provides a pathway for fundamental understanding of phase-transforming flows with multiple applications in science, engineering, and medicine.}


\maketitle

%%-----------------------------------------
%% Introduction
%%-----------------------------------------

Flows of phase-transforming fluids are principal across science, engineering and medicine. Management of electronics cooling, which depends heavily on liquid-vapor flows, remains a critical barrier to creating more powerful datacenter computers and meeting the performance demands of an increasingly computerized society and industry. The collapse of a cavitation bubble, which is another notable example of flows with phase transformations, has fascinated scientists for decades due to the extreme conditions generated, including temperatures of up to 5,000 K, emission of light and strong shock waves and jets \cite{Flint1991-vd}. Although cavitation continues to be a significant concern in the design on marine propellers, it has also been exploited technologically for ultrasonic cleaning and drug delivery \cite{Suslick1999-ce,Suslick1990-ov}. Despite their prevalence and importance, our understanding of fluid flows with phase transformations remains poor, partially due to the challenges they pose to computational methods. Phase-transforming flows involve non-equilibrium thermodynamics, large viscosity and density ratios, moving interfaces with topological changes and flow physics that spans a wide range of time and length scales. The most advanced computational methods are based on compressible flow models for mixtures of liquid and vapor. Although mixture models have been successful in several applications, their approach to phase change is either based on thermodynamic equilibrium or on phenomenological models that enter the mass balance equations and are known to have an important effect on the predictions \cite{Frikha2009-qr}. The latter phase-change models, also called mass-transfer functions, involve parameters that depend on the flow conditions and need frequent re-calibration. Importantly, these models cannot predict nucleation of vapor bubbles from pure liquid, which precludes further mechanistic understanding of, arguably, the most critical problem in cavitating and boiling flows \cite{Brennen2014-xw}.

Interestingly, van der Waals proposed a first-principles thermodynamic theory of liquid-vapor phase change \cite{Van_der_Waals1979-xt}. The model is based on a non-convex Helmholtz bulk free energy extended with a nonlocal term that accounts for interfacial energy. The use of a non-convex bulk thermodynamic potential permits to incorporate state-of-the-art theory of phase transformations that enables the prediction of nucleation and spinodal decomposition. Non-convex potentials have found dramatic success in predicting thermodynamic properties and critical points of liquid-vapor mixtures \cite{kontogeorgis2009thermodynamic}. Importantly, van der Waals' thermodynamic theory can be coupled with the balance equations of compressible flows in a thermodynamically consistent manner that guarantees the second law is satisfied for an arbitrary process compatible with the balance laws. The result of coupling van der Waals theory with flow is the Navier-Stokes-Korteweg (NSK) equations. Although the potential of the NSK equations for mechanistic understanding and prediction of liquid-vapor flows has been exploited to study nucleation \cite{Magaletti2021-qv}, fluid instability under shear \cite{furukawa2006violation} and bubble collapse \cite{Magaletti2015-gi}, current computational methods are limited to micrometer-scale flows without solid walls or flow conditions very close to criticality. Thus, the predictive capability of the NSK equations remains unrealized for a wide range of boiling and cavitating flows at length scales larger than a few micrometers. 

Here, we present unprecedented three-dimensional simulations of wall-bounded cavitating flows at centimeter scale and $\mathcal{O}(10^5)$ Reynolds number using the NSK equations. Because our simulations are based only on van der Waals' thermodynamic theory and fundamental continuum mechanics without additional modeling assumptions, we call them Direct van der Waals simulations (DVS). Our computations are enabled by a new residual-based, stabilized discretization concept that does not require hyperbolicity of the isentropic form of the equations, extends to van der Waals fluids the Streamline Upwind Petrov Galerkin (SUPG) technique \cite{Shakib1991-de,Codoni2021-fi} and the discontinuity capturing operators \cite{Bazilevs2021-uo} and improves the thickened interface methods \cite{Jamet2001-ew,Nayigizente2021-yy}. We illustrate the algorithm's performance with a parametric study of cavitating flow past a cylinder and a simulation of flow over a wedge that shows sheet-to-cloud transition. Our results are in good agreement with experiments, indicating that the proposed algorithm opens the opportunity to predict boiling and cavitating flows at centimeter scale or even larger using minimal modeling assumptions. 

%%-----------------------------------------
%% Results
%%-----------------------------------------

\section{Results}\label{sec:Results}

\subsection{Model Overview} \label{sec:ModelOverview}

The NSK equations are derived from the functional Helmholtz free energy
\begin{equation}
    \label{eqn:Helmholtz}
    %\mathcal{H}(\rho, \nabla \rho)
    {\mathcal{H}[\rho]}=\int_{\Omega} \left( \psi(\rho) + \frac{\lambda\eta}{2}\vert\nabla \rho\vert^2 \right){\rm d}\Omega.
\end{equation}
Here, $\Omega$ is the fluid domain, $\rho$ is the fluid's density, $\psi$ is the bulk Helmholtz free energy per unit volume, while $\lambda$ and $\eta$ are constants that control, respectively, interfacial energy and interface thickness. The thermodynamic potential in \cref{eqn:Helmholtz} differs from standard potentials used for compressible flows in two critical aspects that are interconnected. First, $\mathcal{H}$ depends not only on $\rho$, but also on its gradient. Second, because $\mathcal{H}$ depends on the density gradient, the thermodynamic potential remains convex in the sense of functional derivatives, even if $\psi$ is not; {see Appendix A}. The possibility of utilizing a non-convex bulk free energy per unit volume $\psi$ allows us to use state-of-the-art theory in non-equilibrium phase transformations. From the thermodynamic potential given in \cref{eqn:Helmholtz}, we can derive the NSK equations using balance laws for mass, linear momentum, angular momentum, energy and the second law of thermodynamics. The NSK equations for an isothermal system are
\begin{equation}
\label{eqn:mass_conservation}
    \frac{\partial \rho}{\partial t} + \nabla \cdot (\rho \bm{u}) = 0,
\end{equation}
\begin{equation}
\label{linear_momentum_conservation}
    \frac{\partial (\rho \bm{u})}{\partial t} + \nabla \cdot (\rho \bm{u} \otimes \bm{u} + p \bm{I}) - \nabla \cdot \bm{\tau} - \nabla \cdot \bm{\zeta} = 0,
\end{equation}
where \cref{eqn:mass_conservation,linear_momentum_conservation} represent, respectively, mass and linear momentum conservation. Here, $\bm{u}$ is the fluid velocity, $p=\rho^2 \frac{\partial (\psi / \rho)}{\partial \rho}$ is the fluid pressure and $\bm{I}$ is the identity tensor. The tensor $\bm{\tau}$ denotes viscous stresses, which for a Newtonian fluid under Stokes' hypothesis are given by
\begin{equation}
\label{eqn:viscous}
    \bm{\tau} = \overline{\mu}(\rho)\left(\nabla \bm{u} + \nabla^T \bm{u} -\frac{2}{3}\,\nabla \cdot \bm{u}\, \bm{I}\right),
\end{equation}
where $\overline{\mu}(\rho)$ is the density-dependent viscosity coefficient; see Methods. The Korteweg stress tensor is 
\begin{equation}
\label{eqn:Korteweg}
    \bm{\zeta} = \lambda \eta \left[\left(\rho \Delta \rho + \frac{1}{2}\abs{\nabla \rho}^2\right)\bm{I} - \nabla \rho \otimes \nabla \rho \right],
\end{equation}
and accounts for the interfacial stresses.

The challenges in the simulation of \crefrange{eqn:mass_conservation}{eqn:Korteweg} for wall-bounded, large Reynolds number flows at centimeter scale emanate from two difficulties. First, there is a very large disparity between the length scale at which interfacial physics occurs and the largest length scale that controls flow physics. We address this by proposing the stabilized thickened interface method; see Methods. Second, the inviscid NSK equations with vanishing Korteweg stress are not hyperbolic, which precludes the direct use of most standard computational methods for compressible flows. We bypass this difficulty using residual-based stabilization with shock capturing; see Methods.

To illustrate the potential of DVS, we study cavitating flow over a circular cylinder and over a wedge at centimeter scale. For all cases, we impose free-stream inlet boundary conditions ($\bm{u}_\infty$ and $p_\infty$) using an acoustically absorbing sponge layer \cite{Colonius2004-rv}. The flow conditions are characterized by the free-stream cavitation number and Reynolds number. The free-stream cavitation number is $\sigma_\infty = 2(p_\infty - p_v) / (\rho_\infty u_\infty^2)$, where $\rho_\infty$ is the density that corresponds to $p_\infty$ in our equation of state, and $p_v$ is the vapor pressure. The free-stream Reynolds number is $R_e^\ell=\rho_\infty u_\infty\ell/\overline{\mu}_l$, where $\ell$ is a problem-dependent length scale, and $\overline{\mu}_l$ is the dynamic viscosity in the liquid phase. The strength and extent of cavitation will be measured using the void fraction $\alpha = (\rho_\infty - \rho) / (\rho_\infty - \rho_v)$.

\subsection{Cavitating flow over a circular cylinder}\label{sec:cylinder}

\begin{figure*}[htbp]%htb
  \centering
  \scalebox{0.8}{\includegraphics{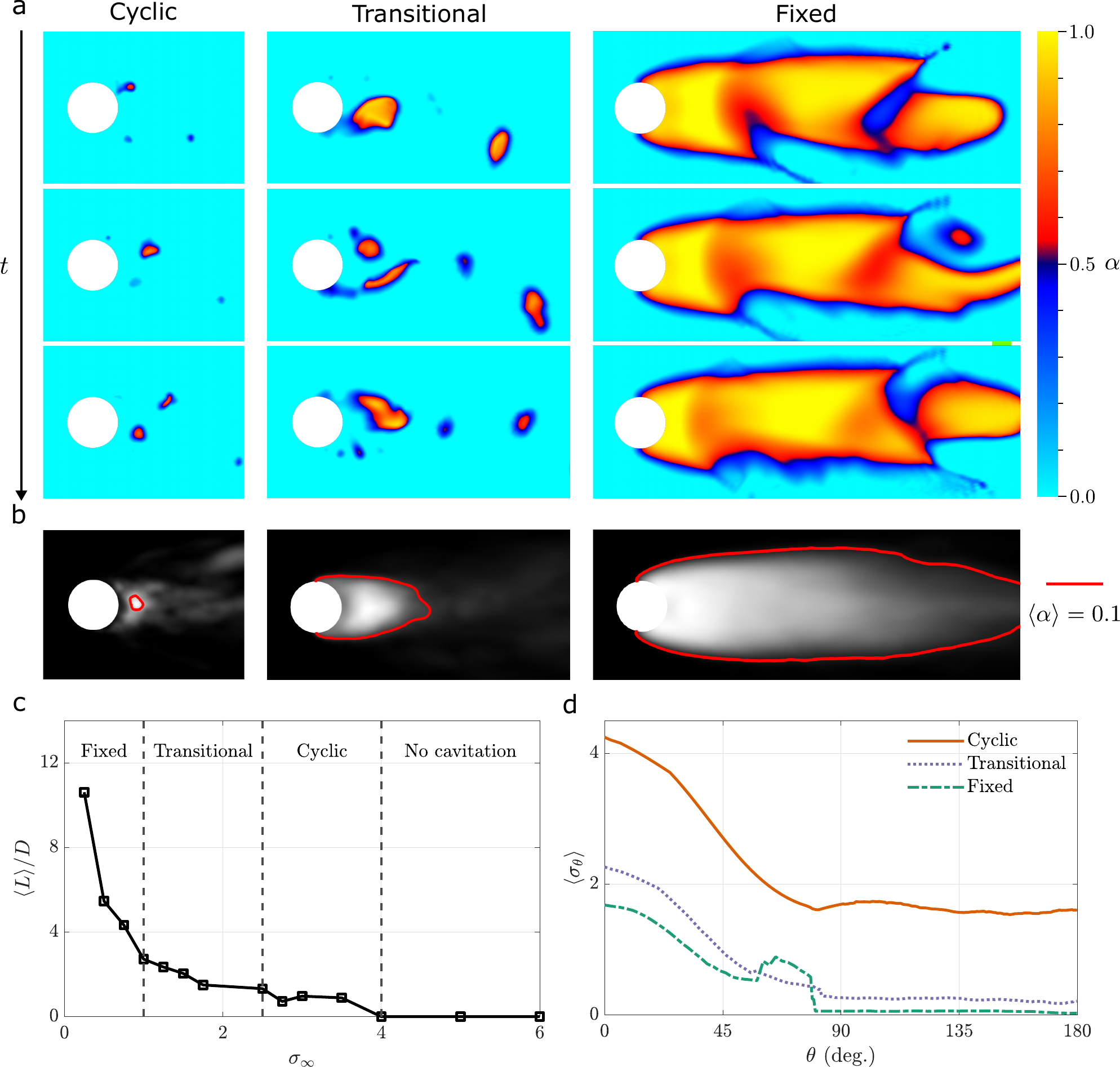}}
  \caption{{Parametric study of cavitating flow over a circular cylinder of diameter $D$. The computations are performed on a two-dimensional domain whose external boundary is an ellipse with a semi-major axis of $30D$ and a semi-minor axis of $11.25D$. The cylinder is located at the center of the ellipse. We use 77,274 $C^1-$continuous quadratic elements to discretize the domain and an acoustically absorbing sponge layer with a width of $2D$ is placed near the edge of the ellipse. The temperature is $T=300$ K. The freestream velocity is $u_\infty=15.2$ m/s, the cylinder's diameter is $D=2$ mm, and the dynamic viscosity of the liquid phase is $\overline{\mu}_l=10^{-3}$ Pa$\cdot$s, which  corresponds to $R_e^D = 2.6\times10^4$. The dynamic viscosity of the vapor phase is $\overline{\mu}_v=10^{-5}$ Pa$\cdot$s; see \cref{eqn:mu_dyn}. We vary the freestream pressure to change the cavitation number. We choose $\lambda = 10^{-16}$ m$^7$/kg/s$^2$ and $\eta = 10^7$ as interfacial parameters, which yields the surface tension for a liquid-vapor interface in water at the problem's length scale. To reduce the computational cost, the initial condition is obtained from an incompressible flow simulation with density $\rho_\infty$, which implies $\alpha=0$}. a. Instantaneous void fraction for free-stream cavitation number $\sigma_\infty$ of 3.0 (Cyclic), 1.25 (Transitional) and 0.25 (Fixed). b. Time-averaged void fraction $\langle\alpha\rangle$. c. Average vapor cavity length non-dimensionalized by cylinder diameter as a function of free-stream cavitation number $\sigma_\infty$. d. Time-averaged local cavitation number $\langle\sigma_\theta\rangle$  distribution on the cylinder.}
  \label{fig:CylinderCavity}
\end{figure*}

\begin{figure*}[htbp]%htb
  \centering
  \scalebox{0.8}{\includegraphics{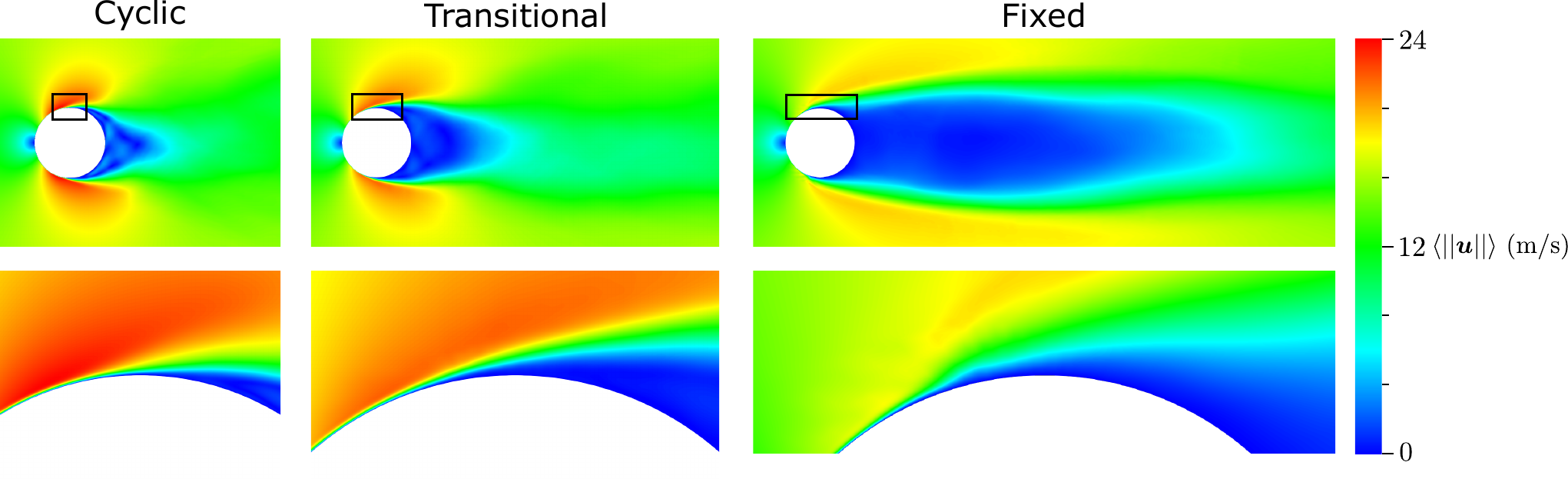}}
  \caption{Parametric study of cavitating flow over a circular cylinder. Time-averaged velocity magnitude on the entire cylinder (top row) and zoomed-in {at the black rectangular regions} (bottom row) for different cavitation modes.}
  \label{fig:CylinderVelocity}
\end{figure*}

A flowing fluid accelerates as it moves around the leading edge of a cylinder. The fluid's acceleration leads to a pressure drop that can trigger cavitation. Flows over cylinders have been often used to study cavitation because, depending on the free-stream conditions, they can feature different types of cavitation and different inception locations. Here, we perform a parametric study varying the free-stream pressure to produce free-stream cavitation numbers that span the range $\sigma_{\infty} = 0.25$ (strong cavitation) to $\sigma_{\infty} = 6.0$ (no cavitation). \Cref{fig:CylinderCavity}a shows snapshots of the instantaneous void fraction for different cavitation numbers under a flow field that goes from left to right. For $\sigma_{\infty}=3$ (left panels) we observe cyclic cavitation. In this cavitation regime, the small cavities formed at the cylinder's surface, detach almost instantaneously and are captured by the vortex immediately downstream of the cylinder. Because of their small sizes, these cavities collapse shortly after leaving the vortex. For transitional cavitation at $\sigma_{\infty}=1.25$ (central column), some vapor pockets separate instantaneously from the cylinder's surface. Some cavities, however, remain attached to the cylinder for a time interval, grow, and eventually are carried downstream by the flow. Our simulation for $\sigma_{\infty}=0.25$ shows fixed cavitation. In this case, a significant fraction of the cylinder surface is consistently covered by vapor. The average cavity length remains stable over time, but its trailing edge continuously sheds gas pockets. The time-averaged vapor fraction $\langle\alpha\rangle$ \ offers a more conclusive picture of the primary location of the cavity for each case; see \Cref{fig:CylinderCavity}b. For cyclic cavitation, the cavity is entirely detached from the cylinder. For transitional cavitation, the time-averaged cavity is attached to the cylinder and has a length that is comparable to the cylinder's diameter. For fixed cavitation, the cavity length is much larger than the cylinder and its thickness also exceeds the cylinder's diameter. \Cref{fig:CylinderCavity}c shows the time-averaged length of the cavity $\langle L \rangle$ relative to the cylinder's diameter as a function of the cavitation number. The results are in good agreement with experiments \cite{Fry1984-vn} and past numerical studies \cite{Gnanaskandan2016-bk}. We observe that although the cavity length decreases monotonically with the cavitation number in the majority of the plot, there is a small region, close to the boundary between transitional and cyclic cavitation where it increases. Interestingly, this counter-intuitive result has also been observed experimentally \cite{Fry1984-vn}. Based on our results, one potential explanation is as follows: as the cavitation regime changes from cyclic to transitional, the size of the cavities attached to the cylinder grows. The presence of larger cavities at the cylinder's surface reduces the vortex strength and leads to weaker cavitation inside the vortex. Although for smaller cavitation number, larger cavities are shed into the free stream, they are short-lived because the free-stream pressure is relatively large and do not contribute significantly to increase \textless$L$\textgreater. Thus, in this regime, the overall effect of the cavitation number increase is a larger cavity length. \Cref{fig:CylinderCavity}d shows the pointwise, time-averaged cavitation number on the cylinder surface, $\langle\sigma_\theta\rangle = 2(\langle p_\theta \rangle - p_v)/(\rho_\infty u_\infty^2)$. Here, $\theta$ is a parametric coordinate along the cylinder's surface such that $\theta=0^{\circ}$ and $\theta=180^{\circ}$ correspond to the leading and trailing edges, respectively. For cyclic cavitation, $\langle\sigma_{\theta}\rangle$ reaches a local minimum at $\theta \approx 80^{\circ}$. For slightly larger values of $\theta$, the pressure first increases due to flow deceleration and later decreases due to cavity shedding. For transitional cavitation, $\langle\sigma_{\theta}\rangle$ decreases monotonically with $\theta$, which reinforces the idea that cavitation inception is caused by instantaneous pressure fluctuations. In contrast with the previous two cases, for fixed cavitation,  $\langle\sigma_{\theta}\rangle$ drops abruptly to zero at $\theta \approx 80^{\circ}$ and remains at this value on the rest of the cylinder's surface. These results further emphasize the difference between the three cavitation modes. Interestingly, we observe that $\langle\sigma_{\theta}\rangle$  has a sharp increase at $\theta \approx 55^{\circ}$ for fixed cavitation. To better understand this phenomenon, we show the time-averaged velocity magnitude for the entire cylinder (top) and near the separation point (bottom) in \cref{fig:CylinderVelocity}. The velocity inside the vapor pocket remains close to zero, which indicates that only a small fraction of the momentum is transported across the liquid-vapor interface. When we have a cavity consistently attached to the cylinder, more kinetic energy accumulates upstream and is converted into internal energy. Such conversion causes a local increment in pressure and a stronger adverse pressure gradient, which leads to the thickening of the boundary layer and earlier flow separation. Such phenomenon has been observed experimentally \cite{Ramamurthy1977-qq, Arakeri1975-gf}, but has remained elusive for computational methods.%} 

\subsection{Sheet-to-cloud transition in cavitating flow over a wedge}\label{sec:wedge}

\begin{figure*}[htbp]%htb
  \centering
  \scalebox{0.8}{\includegraphics{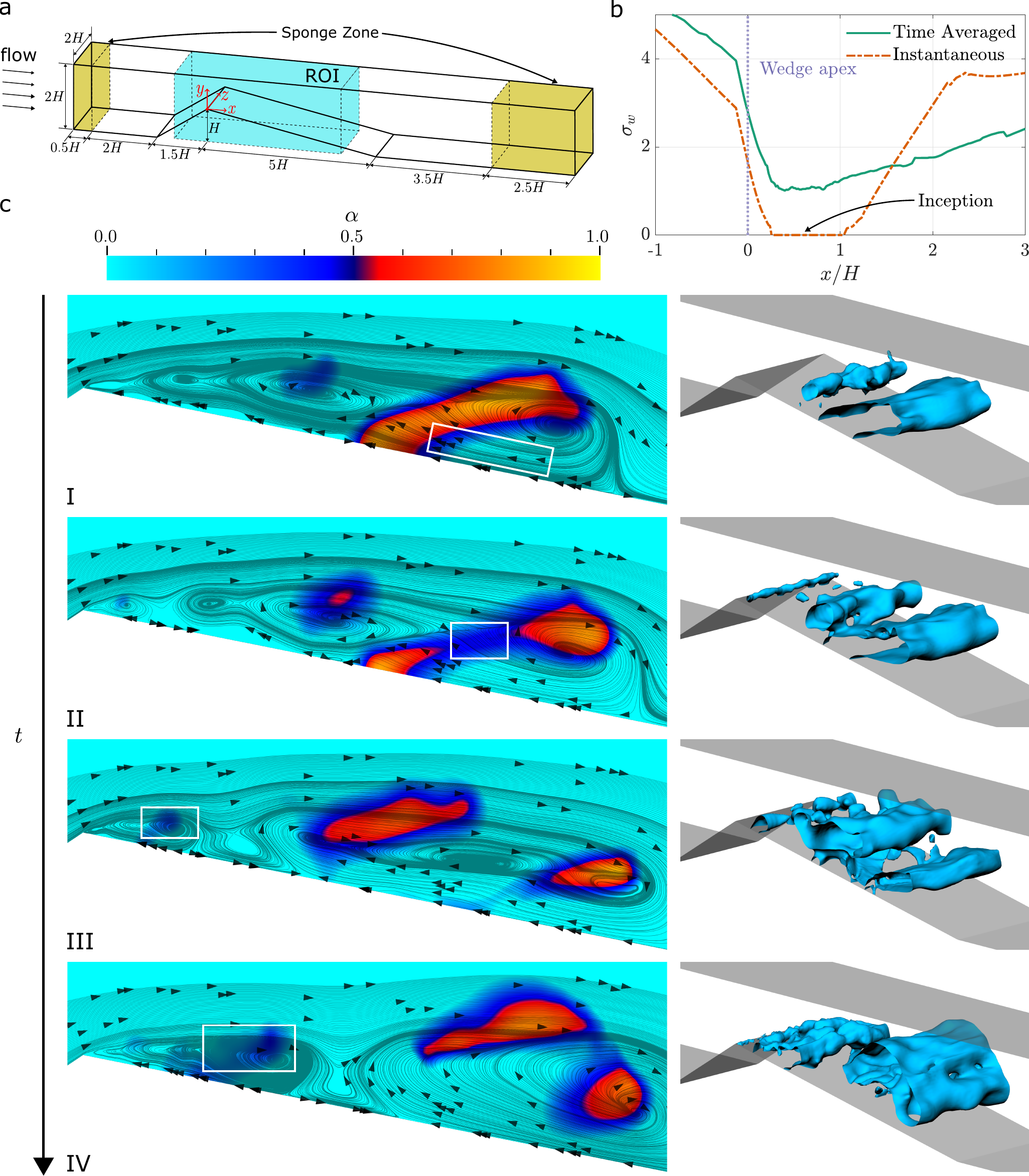}}
  \caption{Sheet-to-cloud transition in cavitation over a wedge. The wedge height is $H = 1.5$ cm, and the flow temperature is $T = 300$ K. The dynamic viscosity of the liquid and vapor phases is, respectively, $\overline{\mu}_l=10^{-3}$ Pa$\cdot$s and $\overline{\mu}_v=10^{-5}$ Pa$\cdot$s; see \cref{eqn:mu_dyn}. The free-stream conditions are $u_\infty = 15.17$ m/s and $p_\infty = 101325$ Pa, which correspond to $R_e^H = 2\times 10^5$ and $\sigma_\infty = 1$. We choose $\lambda = 10^{-16}$ m$^7$/kg/s$^2$ and $\eta = 10^9$ as interfacial parameters, which yields the surface tension for a liquid-vapor interface in water at the problem's length scale. To reduce the computational cost, the initial condition is obtained from an incompressible flow simulation with density $\rho_\infty$, which implies $\alpha=0$. a. Computational domain used for the simulation. The mesh is composed of 542,997 trilinear hexahedral elements. b. Spanwise-averaged  cavitation number along the wedge surface as a function of normalized streamwise distance. c. Instantaneous snapshots of the spanwise-averaged void fraction accompanied by arrowed streamlines (left) and 3D isocontours at void fraction $\alpha=0.2$ (right).}
  \label{fig:Wedge}
\end{figure*}

Flows over a wedge have been often used to study cavitation problems. \Cref{fig:Wedge}a shows a schematic configuration of this physical system and our simulation setup. Under these conditions, the inlet flow accelerates along the wedge, which leads to a pressure drop that triggers cavitation. The cavity initially grows attached to the bottom wall developing the shape of an elongated sheet. The sheet grows longer until it pinches off and transitions to a cloud. The cloud is a three-dimensional structure with features that range across multiple length scales. As the cloud travels downstream, it encounters increasingly large pressures that lead to bubble collapse, which generates jets and sound. The results are in agreement with the experimental observations \cite{Ganesh2016-id}, but reveal important aspects of the cavitation inception process and the sheet-to-cloud transition. Understanding the flow conditions that trigger cavitation remains an outstanding challenge. Our results point to a complex scenario in which cavitation is a strongly unsteady and heterogeneous process that is tightly controlled by localized and instantaneous reductions of pressure. \Cref{fig:Wedge}b shows that the time-averaged pressure remains well above the vapor pressure, but it is instantaneous descents of the pressure, at a level similar to the vapor pressure, that trigger cavitation. \Cref{fig:Wedge}b also illustrates that the instantaneous pressure decreases quickly along the wedge due to flow acceleration, but it does not reach a minimum at the wedge apex. Instead, the boundary layer separation that occurs downstream of the apex generates vortices that undergo stretching and further reduce the pressure, eventually leading to the formation of a vapor cavity. The transition from sheet to cloud cavitation is important because cavitation clouds have a higher potential to generate shock waves and noise. However, the mechanisms that control the transition remain poorly understood \cite{pelz2017transition}. Recent research \cite{bhatt2021cavitating,wu2021cavitation} points to a scenario in which, as the cavity grows, the sheet becomes unstable and transitions into a cloud due to a combination of a re-entrant jet and a condensation shock that travels upstream. Reference \cite{Callenaere2001-gp} identifies two transition types based on the sheet's thickness. In thick sheets, the jet plays a minor role until it reaches the cavity's leading edge and triggers the transition. In contrast, thin cavities break into smaller-scale, three-dimensional structures immediately after they are impinged by the re-entrant jet. \Cref{fig:Wedge}c shows snapshots of the spanwise-averaged instantaneous void fraction. As the sheet cavity travels downstream, the re-entrant jet starts to develop due to the presence of an adverse pressure gradient. Because the sheet is thick, it remains intact as the jet travels through. Once the jet fully penetrates the sheet, a cloud cavity pinches off the rest of the sheet. As the cloud cavity travels downstream, the remaining sheet starts to interact with free stream nuclei and forms a second cloud cavity. Meanwhile, a new and thinner sheet cavity develops near the wedge apex due to pressure fluctuations. While this secondary sheet cavity develops, a new re-entrant jet is formed. Because the secondary sheet cavity is thinner, the re-entrant jet immediately destabilizes it, leading to many smaller-scale three-dimensional structures.
%%

%%-----------------------------------------
%% Discussion
%%-----------------------------------------

\section{Discussion}\label{sec:Discussion}

We propose an algorithm that allows Direct van der Waals simulation (DVS) of phase-transforming fluids for wall-bounded flows far from criticality and large Reynolds numbers at unprecedented length scales. Our algorithm is based on a residual-based formulation and a stabilized thickened interface method. The proposed approach successfully addresses two critical challenges that limited existing computational methods, namely, the non-hyperbolic eigenstructure of the inviscid equations without Korteweg stress and the disparity of length scales between interfacial physics and flow physics. The strength of DVS is that it couples flow dynamics with a fundamental non-equilibrium theory of phase transformations without resorting to phenomenological approaches that require flow-dependent parameter calibration. DVS opens the possibility to gain new mechanistic understanding of the most critical processes of phase-transforming flows, including nucleation of the vapor phase in boiling and cavitation.

To illustrate our approach, we performed a parametric study of flow over a circular cylinder, varying the free-stream pressure. As the free-stream pressure is reduced, DVS predicts a transition from non-cavitating to cavitating flow. DVS also predicts the progression from cyclic to fixed cavitation in quantitative agreement with experiments. Our DVS results indicate that, as the vapor cavity attached to the cylinder's trailing edge grows larger, the separation point moves upstream. This subtle, yet critical phenomenon, has been observed in experiments but not in state-of-the-art cavitation simulations. 

We performed a three-dimensional simulation of cavitating flow over a wedge of 1.5 cm height. Our DVS results capture a highly turbulent flow as well as the transition from sheet to cloud cavitation. The simulation shows that cavitation inception is tightly controlled by local pressure fluctuations and vortex dynamics. In agreement with experiments, DVS shows that thin and thick sheet cavities respond differently to re-entrant jets and condensation shocks, which leads to distinctive destabilization mechanisms of the sheet cavity. Overall, our results highlight the predictive capabilities of DVS, which are particularly noteworthy because the modeling assumptions are minimal. We believe that DVS opens new possibilities not only to simulate and predict flows of phase-transforming fluids, but also to fundamentally understand bubble nucleation and cavitation inception.

%%-----------------------------------------
%% Methods
%%-----------------------------------------

\section{Methods} \label{sec:methods}

%%%%%%%%%%%%%%%%%%%%%%%%%%%%%%%%%%%%%%%%%%%%%

\subsection{Governing equations}\label{sec:GE}
The isothermal Navier-Stokes-Korteweg equations can be written as
\begin{equation}
    \bm{{U}}_{,t} + \bm{{F}}^{\rm adv}_{i,i} = \bm{{F}}^{\rm diff}_{i,i} + \bm{{F}}^{c}.
    \label{eqn:Compact_NonConsForm}
  \end{equation}
Here, an inferior comma denotes partial differentiation (e.g., $\bm{{U}}_{,t}=\partial\bm{U}/\partial t$) and repeated indices indicate summation over the spatial dimensions (e.g., $\bm{{F}}^{\rm adv}_{i,i}=\sum_{i=1}^{d}\partial\bm{{F}}^{\rm adv}_{i}/\partial x_i$, where $x_i$ denotes the $i$th Cartesian coordinate and $d$ is the number of spatial dimensions). The vector $\bm{{U}} = \left[\rho,\rho u_1,\rho u_2,\rho u_3\right]$ contains the conservation variables. The vectors $\bm{{F}}^{\rm adv}_i$, $\bm{{F}}^{\rm diff}_i$ and $\bm{{F}}^{c}_i$ represent, respectively, the advective fluxes, the diffusive fluxes and the Korteweg stress, and they are defined as
\begin{equation}
    \bm{{F}}^{\rm adv}_i = \bm{{F}}^{\rm adv/p}_i + \bm{{F}}^{p}_i = 
    \begin{bmatrix} 
      \rho u_i \\ 
      \rho u_1 u_i\\ 
      \rho u_2 u_i\\
      \rho u_3 u_i
    \end{bmatrix}
    + 
    \begin{bmatrix} 
      0 \\ 
      p\delta_{1i}\\ 
      p\delta_{2i}\\
      p\delta_{3i}
    \end{bmatrix}
\end{equation}
 
\begin{equation}
  \bm{{F}}^{\rm diff}_i = 
  \begin{bmatrix} 
    0 \\ 
    \tau_{1i} \\ 
    \tau_{2i} \\ 
    \tau_{3i} 
  \end{bmatrix}
  \mbox{, }
  \bm{{F}}^{c} = 
  \begin{bmatrix} 
    0 \\ 
    \lambda \eta \rho \Delta \rho_{,1}\\ 
    \lambda \eta \rho \Delta \rho_{,2}\\
    \lambda \eta \rho \Delta \rho_{,3}
  \end{bmatrix}
  \label{eqn:F_U}
\end{equation}
where $\delta_{ij}$ is the Kronecker Delta. In \cref{eqn:F_U}, we have used the identity $\nabla \cdot \bm{\zeta} = \lambda \eta \rho \nabla (\Delta \rho)$. In the viscous stress tensor $\bm{\tau} = \overline{\mu}(\rho)\left(\nabla \bm{u} + \nabla^T \bm{u} -\frac{2}{3}\,\nabla \cdot \bm{u}\, \bm{I}\right)$, the dynamic viscosity is defined as 
\begin{equation}
  \overline{\mu}(\rho) = 
  \begin{cases}
    \overline{\mu}_v,&\mbox{ } 0 <\rho \leq \rho_v \\[1.25em]
    \frac{\rho_l - \rho}{\rho_l - \rho_v}\overline{\mu}_v + \frac{\rho - \rho_v}{\rho_l - \rho_v}\overline{\mu}_l,&\mbox{ } \rho_v < \rho < \rho_l  \\[1.25em]
    \overline{\mu}_l,&\mbox{ } \rho_l\leq\rho,
  \end{cases}
  \label{eqn:mu_dyn}
\end{equation}
where $\overline{\mu}_{l/v}$ and $\rho_{l/v}$ are, respectively, the dynamic viscosity and saturation density for the liquid and vapor phases. To derive our algorithm, we define the primitive variables $\bm{Y} = \left[\rho, u_1, u_2, u_3\right]$ and the following transformation matrices,
\begin{alignat}{2}
  %\begin{split}
    &\bm{A}_0 = \frac{\partial \bm{{U}}}{\partial \bm{Y}} \mbox{, } \bm{{A}}_i^{\rm adv/p} = \frac{\partial \bm{{F}}^{\rm adv/p}_i}{\partial \bm{Y}} \mbox{, } \bm{{A}}_i^{p} = \frac{\partial \bm{{F}}^{p}_i}{\partial \bm{Y}} \\
    &\bm{{A}}_i^{\rm adv} = \bm{{A}}_i^{\rm adv/p} + \bm{{A}}_i^{p} \\
    &\bm{A}^c_i \bm{Y}_{,i}= { \bm{F}^c}{}\\
    &\bm{{K}}_{ij}{\bm{Y}_{,j}} = {\bm{{F}}_i^{\rm diff}}
  %\end{split}
\end{alignat}
whose explicit expressions are given in Appendix B. Using the transformation matrices we can rewrite \cref{eqn:Compact_NonConsForm} in quasi-linear form
\begin{equation}
  \bm{{A}}_0 \bm{Y}_{,t} + \bm{{A}}_i^{\rm adv/p} \bm{Y}_{,i} + \bm{{A}}_i^{p} \bm{Y}_{,i} = \left(\bm{{K}}_{ij} \bm{Y}_{,j}\right)_{,{\color{black}{i}}} + \bm{A}_i^{c}\bm{Y}_{,i}
  \label{eqn:LocalQuasiLinearForm}
\end{equation}

\subsection{Cubic equation of state} \label{sec:EoS}

Cubic equations of state (EoS) are widely used to represent liquid-vapor equilibrium \cite{kontogeorgis2009thermodynamic}. The first cubic EoS is due to van der Waals \cite{Van_der_Waals1979-xt}, but many variants and extensions have been proposed thereafter, including the Soave-Redlich-Kwong (SRK) \cite{Soave1972-zh} and Peng-Robinson models \cite{Peng1976-lw}. Here, we use the EoS
\begin{equation}
    p^{\rm EoS} (\rho, T) = R b \frac{\rho T}{b - \rho} - a(T)b^2 \frac{\rho^2}{b^2 + 2\rho b - \rho^2},
    \label{eqn:EoS_PRSV2}
\end{equation}
where $R$ is the specific gas constant, $T$ is the temperature which is a constant for isothermal conditions, and $a(T)$ and $b$ depend on the fluid. \Cref{eqn:EoS_PRSV2} was proposed in \cite{Stryjek1986-zw}, and provides accurate predictions for liquid-vapor mixtures. For water, the parameter values are $R = 461.5$ J/kg$\cdot$K, and $b = 949.7$ kg/m$^3$. The value of $a(T)$ in units Pa$\cdot$m$^6$/kg$^2$ is 
\begin{equation}
a(T) = 1848.2\left(1 + k(T_R)(1-\sqrt{T_R})\right)^2,
\end{equation}
where $T_R = T/T_c$, the critical temperature is $T_c = 647.1$ K and 
\begin{alignat}{2}
  &k(T_R) = \left[-0.066 + 0.02\left(0.44 - T_R\right)(1 - \sqrt{T_R})\right] \nonumber \\
  &\times (1 + \sqrt{T_R})\left(0.7-T_R\right) + 0.87.
\end{alignat}

%%%%%%%%%%%%%%%%%%%%%%%%%%%%%%%%%%%%%%%%%%%%%

\subsection{Stabilized thickened interface method (sTIM)} \label{sec:sTIM}

\begin{figure}[htbp]%htb
     \centering
     \scalebox{0.35}{\includegraphics{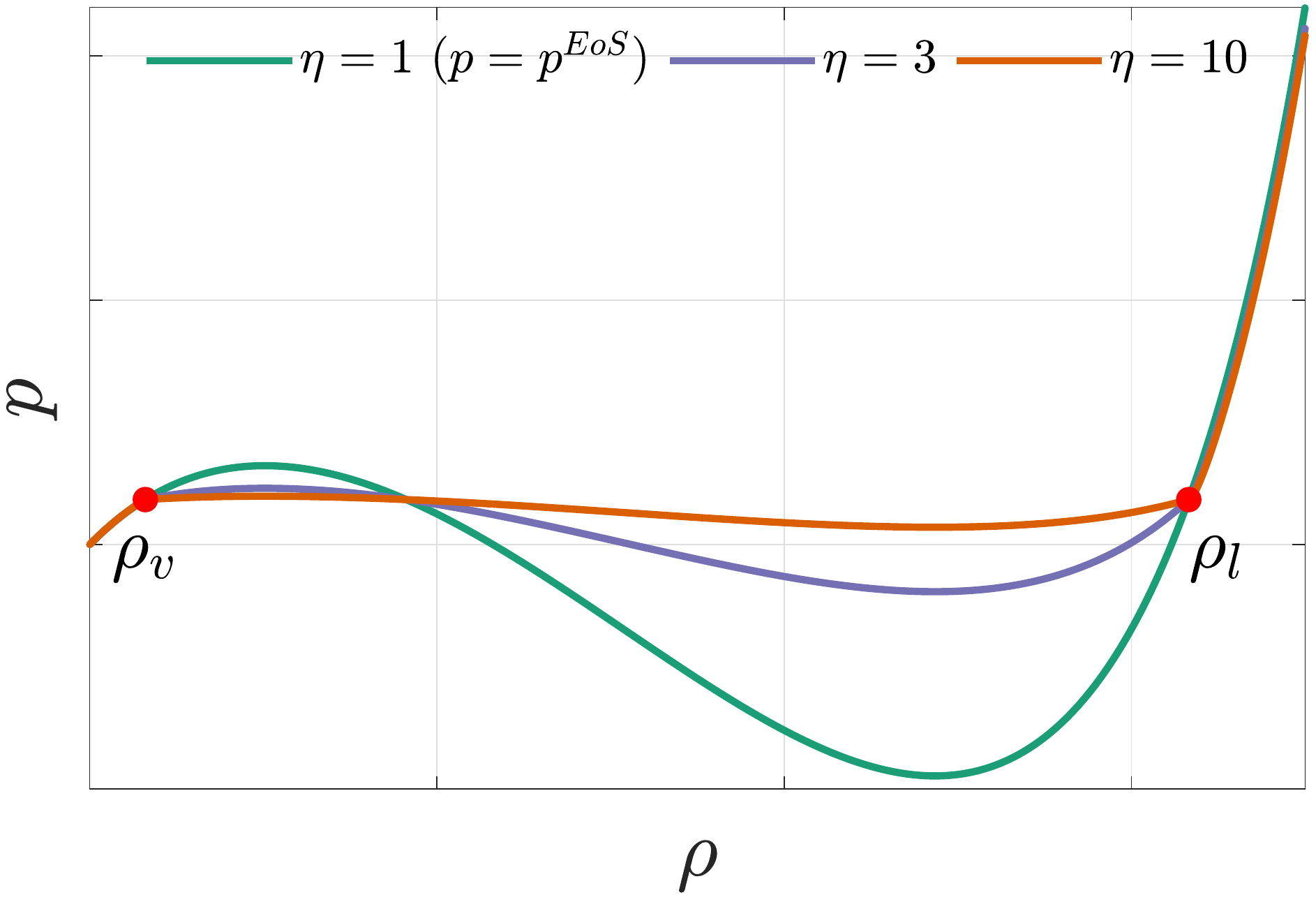}}
     \caption{Comparison of $p^{\rm EoS}$ [\cref{eqn:EoS_PRSV2}] and $p$ [\cref{eqn:p_sTIM}] for $\xi = 0.01$ and several values of $\eta$. For $\eta=1$, $p=p^{\rm EoS}$. For increasing values of $\eta$, $p$ becomes closer to the saturation pressure in the interfacial region $\rho\in(\rho_v,\rho_l)$, while remaining nearly identical to $p^{\rm EoS}$ outside of the interface. This shows that the proposed approach sTIM effectively allows to increase the problem's length scale, while maintaining the thermodynamic properties of the bulk phases.}
     \label{fig:sTIM}
\end{figure}

For temperatures below the critical temperature, equilibrium solutions of the NSK equations with \cref{eqn:EoS_PRSV2} predict a liquid-vapor interface described by a continuous variation of density. At room temperature, the model predicts an interface thickness of less than $100$ nm, in agreement with experiments and molecular dynamics simulations \cite{Yang2020-kc,Dang1997-kk}. In a simulation of the NSK equations, the interface thickness needs to be resolved by the computational mesh which implies that a three-dimensional centimeter-scale computation would require at least $\sim 10^{16}$ degrees of freedom which is prohibitive in today's computer architectures. Enlargement of the interface can be achieved by increasing the parameter $\eta$ in the governing equations. However, increasing $\eta$ without modifying the EoS leads to an overprediction of surface tension that would make the results invalid. Notably, the use of the thickened interface method \cite{Jamet2001-ew,Nayigizente2021-yy} permits to enlarge the interface thickness, while keeping surface tension constant. This is accomplished by increasing $\eta$ and modifying accordingly the binodal region of the EoS. While the thickened interface method opens the possibility to perform larger-scale computations, it leads to the use of a non-differentiable EoS. The lack of smoothness in the EoS leads to the formation of strong spurious shock waves at the interface that propagate throughout the computational domain and become a source of instability. To address this issue, we propose the stabilized thickened interface method (sTIM). The formulation of sTIM is 
\begin{equation}
  p = 
  \begin{cases}
    p^{\rm EoS} + \frac{A_v \xi \rho_v\rho}{(1+\xi)\rho_v - \rho} - A_v \rho,&\mbox{ } 0 <\rho \leq \rho_v \\[1.25em]
    p^{\rm sat} + \frac{p^{\rm EoS}(\rho)-p^{\rm sat}}{\eta},&\mbox{ } \rho_v < \rho < \rho_l  \\[1.25em]
    p^{\rm EoS} + \frac{A_l \xi\rho_l\rho}{(1-\xi)\rho_l - \rho} + A_l \rho,&\mbox{ } \rho_l\leq\rho,
  \end{cases}
  \label{eqn:p_sTIM}
\end{equation}
where $A_{v/l} = \xi\frac{1-\eta}{\eta} \pfrac{p^{\rm EoS}}{\rho} (\rho_{v/l}, T)$, and $\rho_v$, $\rho_l$ are, respectively the vapor and liquid saturation densities at temperature $T$. \Cref{eqn:p_sTIM} shows that in the binodal region, $\rho_v < \rho < \rho_l$, the pressure is modified using the approach proposed in \cite{Nayigizente2021-yy} and, thus, by increasing $\eta$, one can enlarge the interface while keeping surface tension constant. In the vapor ($0 < \rho \le \rho_v$) and liquid ($\rho \ge \rho_l$) phases the original EoS is modified with a stabilizing term whose strength is controlled by the parameter $\xi$. The stabilizing term is designed such that the following conditions are satisfied.
\begin{alignat}{2}
    &p(\rho_v,T) = p^{\rm sat}(T), \label{eqn:p_sTIM_rhov} \\
    &p(\rho_l,T) = p^{\rm sat}(T), \label{eqn:p_sTIM_rhol} \\
    &\pfrac{p}{\rho}(\rho_v,T) = \frac{1}{\eta}\pfrac{p^{\rm EoS}}{\rho}(\rho_v,T), \label{eqn:diff_p_sTIM_rhov} \\
    &\pfrac{p}{\rho}(\rho_l,T) = \frac{1}{\eta}\pfrac{p^{\rm EoS}}{\rho}(\rho_l,T). \label{eqn:diff_p_sTIM_rhol}
\end{alignat}
\Cref{eqn:p_sTIM_rhov,eqn:p_sTIM_rhol} guarantee that saturation pressure remains unchanged, while \cref{eqn:diff_p_sTIM_rhov,eqn:diff_p_sTIM_rhol} ensure that the pressure is a differentiable function at saturation conditions for all $\xi \ne 0$. When the stabilizing term is absent ($\xi  = 0$), the sTIM reduces to the methodology proposed in \cite{Nayigizente2021-yy}. In our computations, we took $\xi = 0.01$, which guarantees that the EoS is smooth and produces changes in the pressure outside of the binodal region that are negligible. \Cref{fig:sTIM} shows a plot of $p$ and $p^{\rm EoS}$ as functions of the density for $\xi=0.01$ and several values of $\eta$. We can see that outside of the binodal region $p$ is indistinguishable from $p^{\rm EoS}$. In the binodal region, $p$ is different from $p^{\rm EoS}$ for $\eta \ne 1$ to achieve the desired effect of decoupled interface thickness and surface tension. The larger is $\eta$, the flatter is $p$ in the binodal region.
%%

%%%%%%%%%%%%%%%%%%%%%%%%%%%%%%%%%%%%%%%%%%%%%
 
\subsection{Variational operators}
\subsubsection{Galerkin operator}
The proposed computational method is based on a weak form of the NSK equations that is stabilized with residual-based terms. Our weak formulation makes use of several semilinear forms. The first one, which emanates from the weak form of \cref{eqn:LocalQuasiLinearForm} without stabilizing terms, is defined as
\begin{alignat}{2}
  %\begin{split}
    &\bm{B}_{\rm NSK}(\bm{W},\bm{Y}) =\nonumber\\
    &\int _\Omega \bm{W} \cdot \left(\bm{{A}}_0 \bm{Y}_{,t} + \bm{{A}}_i^{\rm adv/p} \bm{Y}_{,i} - \bm{A}_i^{c}\bm{Y}_{,i}\right) {\rm d} \Omega \nonumber\\
     - & \int_\Omega \bm{W}_{,i} \cdot \left(\bm{{F}}^{p}_i - \bm{{F}}_i^{\rm diff}\right) {\rm d} \Omega\nonumber\\
     + &\int _\Gamma \bm{W} \cdot \left(\bm{{F}}^{p}_i - \bm{{F}}_i^{\rm diff}\right) n_i {\rm d} \Gamma.
  %\end{split}
  \label{eqn:WeakForm}
\end{alignat}
where $\bm{W} \in {V}$ is a vector-valued weight function, ${V}$ is a suitably chosen functional space, $\Omega$ is the computational domain, $\Gamma$ is the boundary of $\Omega$ and $n_i$ is the $i$th cartesian coordinate of the unit outward normal to $\Gamma$. 

%%%%%%%%%%%%%%%%%%%%%%%%%%%%%%%%%%%%%%%%%%%%%

\subsubsection{SUPG operator} \label{sec:SUPG}
Streamline-Upwind/Petrov-Galerkin (SUPG) is a finite element stabilization method for advection-dominated flow that is applicable to incompressible and compressible flows \cite{Brooks1982-og,Hughes1984-gx}. SUPG is a residual-based stabilizing scheme that provides stable solutions retaining optimal rate of convergence. Let us assume that the domain $\Omega$ is divided into $N_{el}$ elements each denoted by $\Omega^e$. We define the SUPG operator as 
\begin{alignat}{2}
  &\bm{B}_{\rm SUPG}\left(\bm{W},\bm{Y}\right) = \nonumber \\
  &\sum ^ {N_{el}} _ {e=1} \int _{\Omega^e} \left(\bm{A}_i^{*T} \bm{W}_{,i}\right) \cdot \bm{\tau}_{\rm SUPG} {\bf Res}(\bm{Y}) {\rm d} \Omega,
    \label{eqn:SUPG}
\end{alignat}
Here, 
\begin{alignat}{2}
{\bf Res}(\bm{Y}) & =  \bm{{A}}_0 \bm{Y}_{,t} + \bm{{A}}_i^{\rm adv/p} \bm{Y}_{,i} + \bm{{A}}_i^{p} \bm{Y}_{,i} \nonumber\\
                  & -  \left(\bm{{K}}_{ij} \bm{Y}_{,j}\right)_{,i} - \bm{A}_i^{c}\bm{Y}_{,i},
\end{alignat}
is the residual of the governing equations, $\bm{\tau}_{\rm SUPG} = \bm{{A}}_0 ^{-1} \bm{\hat{\tau}}_{\rm SUPG}$ is the stabilizing matrix for the primitive variables and $\bm{\hat{\tau}}_{\rm SUPG}$ is the stabilizing matrix for the conservation variables, which is defined as \cite{Shakib1991-de,Codoni2021-fi}
\begin{equation}
  \begin{split}
  &\bm{\hat{\tau}}_{\rm SUPG} = \\
  &\left(\frac{4\bm{I}}{\Delta t^2}  + G_{ij}\bm{\hat{A}}^{*}_i \bm{\hat{A}}^{*}_j + C_I G_{ij} G_{kl} \bm{\hat{K}}_{ik} \bm{\hat{K}}_{jl}\right)^{-\frac{1}{2}},
  \label{eqn:tau_hat_SUPG}
  \end{split}
\end{equation}
In \cref{eqn:tau_hat_SUPG}, $\Delta t$ is the time step size, $C_I$ is a positive constant derived from an element-wise inverse estimate \cite{Johnson2012-ms}, and $G_{ij}$ represents the components of the element metric tensor $\bm{G}$, that is,
\begin{equation}
  G_{ij} =  \pfrac{\Xi_k}{x_i}\pfrac{\Xi_k}{x_j},
  \label{eqn:GeometricTensor}
\end{equation}
where %$d$ is problem dimension and 
$\bm{x}(\bm{\Xi})$ is the element isoparametric mapping. The matrices $\bm{\hat{A}}^{*}_i$ and $\bm{\hat{K}}_{ij}$ are the conservation variable counterpart of $\bm{{A}}^{*}_i$ and $\bm{{K}}_{ij}$, which can be obtained as,
\begin{equation}
  \bm{\hat{K}}_{ij} = \bm{{K}}_{ij} \bm{{A}}_0^{-1} \mbox{, }\bm{\hat{A}}^{*}_i = \bm{{A}}^{*}_i \bm{{A}}_0^{-1}.
  \label{eqn:K_A_conserve}
\end{equation}
Because the SUPG operator defined in \cref{eqn:SUPG} is residual based, the matrices $\bm{A}^*_i$ can be chosen in multiple ways without compromising the accuracy of the algorithm. However, a poor choice of $\bm{A}^*_i$ will have a detrimental effect on the stability of the scheme. In classical gas dynamics, this choice is guided by an eigenvalue analysis of the isentropic system. Importantly, the isentropic NSK equations are not hyperbolic on the entire phase space and the eigenvalue analysis cannot be used. Based on scaling arguments and local equilibrium at the liquid-vapor interface we choose
\begin{equation}
\label{eqn:Astar}
    \bm{{A}}^{*}_i =  \bm{{A}}^{\rm adv/p}_i + \bm{{A}}^{p}_i - \bm{A}^{c,eq}_i,
\end{equation}
where $\bm{A}^{c,eq}_i$ is an approximation to $\bm{A}^c_i$. In \crefrange{eqn:p_sTIM_rhov}{eqn:diff_p_sTIM_rhol}, $\bm{A}^c_i$ takes on the form
\begin{equation}
    \bm{A}^c_i = \frac{\lambda \eta \rho \Delta \rho_{,i}}{\rho_{,i}} \bm{e}_{i+1} \otimes \bm{e}_{i}\text{ (no sum on $i$)},
\end{equation}
where $\bm{e}_i$ is the $i$th vector of the Cartesian basis in dimension $d+1$. The computation of $\bm{A}^c_i$ is ill-conditioned, especially in the bulk phases. Using $\bm{A}^c_i$ to compute the $\bm{A}^*_i$ matrices leads to small perturbations in the numerical solution that are eventually amplified unless the time step is extremely small. To derive an approximation to $\bm{A}^c_i$, we proceed as follows: under equilibrium conditions, the equation $p_{,i} - \lambda \eta \rho \Delta \rho_{,i} = 0$ is satisfied. Doing basic manipulations, one can show that $p_{,\rho} = \lambda \eta \rho \Delta \rho_{,i} / \rho_{,i}$, where no sum on $i$ is implied. In addition, we know that under equilibrium $p_{,\rho} \le 0$ in the interfacial region and $p_{,\rho} \approx 0$ in the bulk phase. Thus we define
\begin{equation}
    \bm{A}^{c,eq}_i =  p_{,\rho}^{-} \bm{e}_{i+1} \otimes \bm{e}_i, \text{ (no sum on $i$)}
\end{equation}
where $p_{,\rho}^{-}=\min(0, p_{,\rho})$. Importantly, although our derivation of the matrices $\bm{A}^*_i$ assumes that the interface is under local equilibrium conditions, this assumption does not compromise the accuracy of the algorithm in any way. Our discretization method still features high-order accuracy because the $\bm{A}^*_i$ matrices are used only in the SUPG operator which also involves the residual.

This completes the definition of all the matrices on the right-hand side of \cref{eqn:tau_hat_SUPG}. To calculate $\hat{\bm\tau}_{\rm SUPG}$ we need to compute the square root of a $(d+1) \times (d+1)$ matrix. In our simulations, this is done using the Denman-Beavers algorithm \cite{Denman1976-nm,Xu2017-gq}.

%%%%%%%%%%%%%%%%%%%%%%%%%%%%%%%%%%%%%%%%%%%%%%%%

\subsubsection{Discontinuity capturing} \label{sec:DC}

While the use of SUPG ensures stability and accuracy when the solution is smooth, it does not resolve effectively flow fields with shock waves \cite{Le_Beau1993-fp}. We address this by adding to the formulation a residual-based discontinuity-capturing (DC) operator; see \cite{Hughes1986-vj,Hughes1986-ps}. The DC operator for primitive variables is given by
\begin{equation}
    \bm{B}_{\rm DC} \left(\bm{W},\bm{Y}\right) = \sum ^ {N_{el}}_{e=1} \bm{W}_{,i} \cdot \hat{\bm{\kappa}}_{\rm DC} {\bm{A}}_0 \bm{Y}_{,i} {\rm d}\Omega.
  \label{eqn:DC}
\end{equation}
Here, $\hat{\bm{\kappa}}_{\rm DC} = \hat{\kappa}_{C} \bm{e}_i \otimes \bm{e}_i + \hat{\kappa}_M \bm{e}_{i+1} \otimes \bm{e}_{i+1}$ is a $(d+1) \times (d+1)$ diagonal matrix with entries
\begin{equation}
    \hat{\kappa}_c = \min \left( C_C \hat{\kappa}, \hat{\kappa}_{\rm cap} \right)
\end{equation}
\begin{equation}
    \hat{\kappa}_M = \min \left( C_M \hat{\kappa}, \hat{\kappa}_{\rm cap} \right)
\end{equation}
where
\begin{equation}
    \hat{\kappa} = \beta \frac{\vert p_{,\rho}\vert \vert {\bf Res}_1(\bm{Y})\vert + \norm{\bm{u}} \norm{{\bf Res}_{2:d+1}(\bm{Y})}} {\left(\vert p_{,\rho}\vert \nabla {\bm{U}}_1 \otimes \nabla {\bm{U}}_1 + \norm{\bm{u}}\nabla {\bm{U}}_i \otimes \nabla {\bm{U}}_i\right): \bm{G}}
  \label{eqn:kappa_DC_hat}
\end{equation}
\begin{alignat}{2}
  %\begin{split}
  \hat{\kappa}_{\rm cap} &= \beta \bigl( \bm{u}_{\rm rel}\otimes\bm{u}_{\rm rel} : \bm{G^{-1}}
  & + %\mbox{max}\left(p_{,\rho},0\right)
  p_{,\rho}^{+}\mbox{tr}\left(\bm{G}^{-1}\right)\bigr)^{1/2}
  %\end{split}
  \label{eqn:kappa_DC_cap}
\end{alignat}
and $C_C$, $C_M$ are $\mathcal{O}(1)$ positive constants for which we used the value $C_C = C_M = 0.1$. In \cref{eqn:kappa_DC_hat,eqn:kappa_DC_cap}, $\vert{\bf Res}_1(\bm{Y})\vert$ is the absolute value of the residual of the mass conservation equations, $\norm{\cdot}$ denotes the Euclidean norm of a vector, $\bm{\mbox{Res}}_{2:d+1}(\bm{Y})$ is the residual of the linear momentum balance equation, $p_{,\rho}^{+}= \mbox{max}\left(p_{,\rho},0\right)$ and $\bm{u}_{\rm rel }=\bm{u}-\bm{u}_{\infty}$ is the relative velocity with respect to the free-stream velocity. The scaling term $\beta$ is designed to minimize numerical dissipation in the liquid phase while retaining stability in the vapor phase. We use the expression
\begin{equation}
  \beta = 
  \begin{cases}
    \min \left(\beta_{\max},\rho_m / \rho\right),& \rho \leq \rho_m \\%[1.5em]
    1,&\rho_m < \rho \leq \rho_v  \\%[1.5em]
    (\rho_l - \rho)/(\rho_l - \rho_v),& \rho_v < \rho < \rho_l  \\%[1.5em]
    0,&\rho_l \leq \rho.
  \end{cases}
  \label{eqn:alpha_DC}
\end{equation}
where $\rho_m$ represents a very small value of the density and $\beta_{\rm max}$ is a constant that sets the maximum strength of the DC operator. In our simulations, we take $\rho_m = 0.01$ kg/m$^3$ and $\beta_{\max} = 1000$. In the liquid phase, the speed of sound is high, the solution is primarily smooth, and the use of the DC operator is not necessary. In the interfacial region ($\rho_v < \rho < \rho_l$), the value of $\beta$ varies linearly between zero and one. In the vapor phase ($\rho \le \rho_v$), the fluid is highly compressible and the use of a robust DC is necessary to retain numerical stability. For densities in the range $\rho_m < \rho \le \rho_v$, we set $\beta = 1$, which is a commonly used value in gas dynamics simulations \cite{Bazilevs2021-uo}. The DC is maximum when $\rho \le \rho_m$ to avoid the appearance of negative densities. 

%%%%%%%%%%%%%%%%%%%%%%%%%%%%%%%%%%%%%%%%%%%%%%%%%

\subsection{Fully discrete formulation}

The NSK equations include third-order derivatives of the density. Thus, for the operators introduced in \cref{eqn:WeakForm} to be well defined, we need a discrete functional space that is  at least globally $C^1$-continuous. Although we employ a spatial discretization based on Isogeometric Analysis (IGA) that offers this capability \cite{Hughes2005-ct}, classical finite elements do not support globally $C^1$-continuous spaces on complex three-dimensional geometries. Thus, to make our algorithm applicable to classical finite elements we use the split approach which is based on introducing the additional unknown:
\begin{equation}
    \mu = \lambda \eta \Delta \rho.
\end{equation}
By treating $\mu$ as an independent unknown, we can redefine the matrix $\bm{A}^c_i$ as 
\begin{equation}
    \bm{A}^{c,s}_i = \frac{\mu_{,i}}{\rho_{,i}} \bm{e}_{i+1} \otimes \bm{e}_i, \text{ (no sum on $i$)} 
\end{equation}
and rewrite the NSK equations as a larger system of equations with derivatives of order less or equal than two. To formulate our semi-discrete problem we use a finite element space $V^h$ that satisfies the Dirichlet boundary conditions and $V^h_0$, an analogous discrete space that satisfies homogeneous conditions at the Dirichlet boundary. The semi-discretized problem is: find $\left\{ \bm{Y}^h, \mu^h \right \} \in V^h$, such that for all $\left\{ \bm{W}^h,Q^h \right \} \in V^h_0$
\begin{equation}
    \bm{B}_{\rm DVS} \left ( \{\bm{W}^h, Q^h\} \{\bm{Y}^h, \mu^h\} \right ) = 0,
\end{equation}
where 
\begin{alignat}{2}
%\begin{aligned}
    &\bm{B}_{\rm DVS}\left ( \{\bm{W}^h, Q^h\} \{\bm{Y}^h, \mu^h\} \right ) = \nonumber\\
    &\int _\Omega \bm{W}^h \cdot \left(\bm{{A}}_0 \bm{Y}^h_{,t} + \bm{{A}}_i^{\rm adv/p} \bm{Y}^h_{,i} - \bm{A}_i^{c,s}\bm{Y}^h_{,i}\right) {\rm d} \Omega \nonumber\\
    & - \int_\Omega \bm{W}^h_{,i} \cdot \left(\bm{{F}}^{p}_i - \bm{{F}}_i^{\rm diff}\right) {\rm d} \Omega \nonumber\\
    &+ \int _\Omega \left( Q^h \mu^h + Q^h_{,i} \lambda \eta \rho^h_{,i} \right) {\rm d}\Omega \nonumber\\
    &+\sum ^ {N_{el}} _ {e=1} \int _{\Omega^e} \left(\bm{{A}}_i^{*T} \bm{W}^h_{,i}\right) \cdot \bm{\tau}_{\rm SUPG} {\bf Res}(\bm{Y}^h,\mu^h) {\rm d} \Omega\nonumber\\
    &+\sum ^ {N_{el}}_{e=1} \bm{W}^h_{,i} \cdot \hat{\bm{\kappa}}_{\rm DC} {\bm{A}}_0 \bm{Y}^h_{,i} {\rm d}\Omega\nonumber\\
    & + \int _\Gamma \bm{W}^h \cdot \left(\bm{{F}}^{p}_i - \bm{{F}}_i^{\rm diff}\right) n_i {\rm d} \Gamma \nonumber\\
    &- \int _\Gamma Q^h \lambda \eta \rho^h_{,i}  n_i {\rm d} \Gamma.
%\end{aligned}
\end{alignat}
We used the generalized-$\alpha$ method to perform time integration \cite{Jansen2000-gg}. At each time step, the nonlinear system of equations is solved using Newton-Raphson's method with a relative tolerance of $2.5\times10^{-4}$. The linear systems of equations are solved using the GMRES method \cite{Saad1986-ap} with an additive Schwarz preconditioner. The time step is varied throughout the simulation to achieve convergence of the Newton-Raphson algorithm in 3 to 4 iterations. Our code makes use of the open-source package PETSc \cite{Balay2021-vc} and PetIGA \cite{Dalcin2016-cg}.

%%%%%%%%%%%%%%%%%%%%%%%%%%%%%%%%%%%%%%%%%%%%
\subsection{Stability and accuracy of the proposed algorithm}

Because the SUPG operator in \cref{eqn:SUPG} vanishes when the residual $ {\bf Res}(\bm{Y})$ is zero, the matrices $\bm{A}_i^*$ in \cref{eqn:SUPG} can be chosen in multiple ways without compromising the rate of convergence of the algorithm. However, a poor choice of $\bm{A}_i^*$ can make the algorithm unstable for mesh sizes or time steps that are not sufficiently small to reach the asymptotic regime of the algorithm. Here, we show that two choices of $\bm{A}_i^*$ that are logical extensions of the matrices used in standard gas dynamics simulations render unsatisfactory results, while our choice, given by \cref{eqn:Astar}, produces vastly superior results. The alternatives to \cref{eqn:Astar} that we study here are: (I) $\bm{A}_i^*=\bm{A}_i^{\rm adv/p}+\bm{A}_i^p$, and (II) $\bm{A}_i^*=\bm{A}_i^{\rm adv/p}+\bm{A}_i^p-\bm{A}_i^c$. Case I corresponds to the standard SUPG operator used for compressible Navier-Stokes; see \cite{Codoni2021-fi}. Case II represents a plausible, but unsuccessful, extension of the SUPG operators from compressible Navier-Stokes to the NSK equations.

\begin{figure*}[htbp]%htb
  \centering
  \scalebox{0.8}{\includegraphics{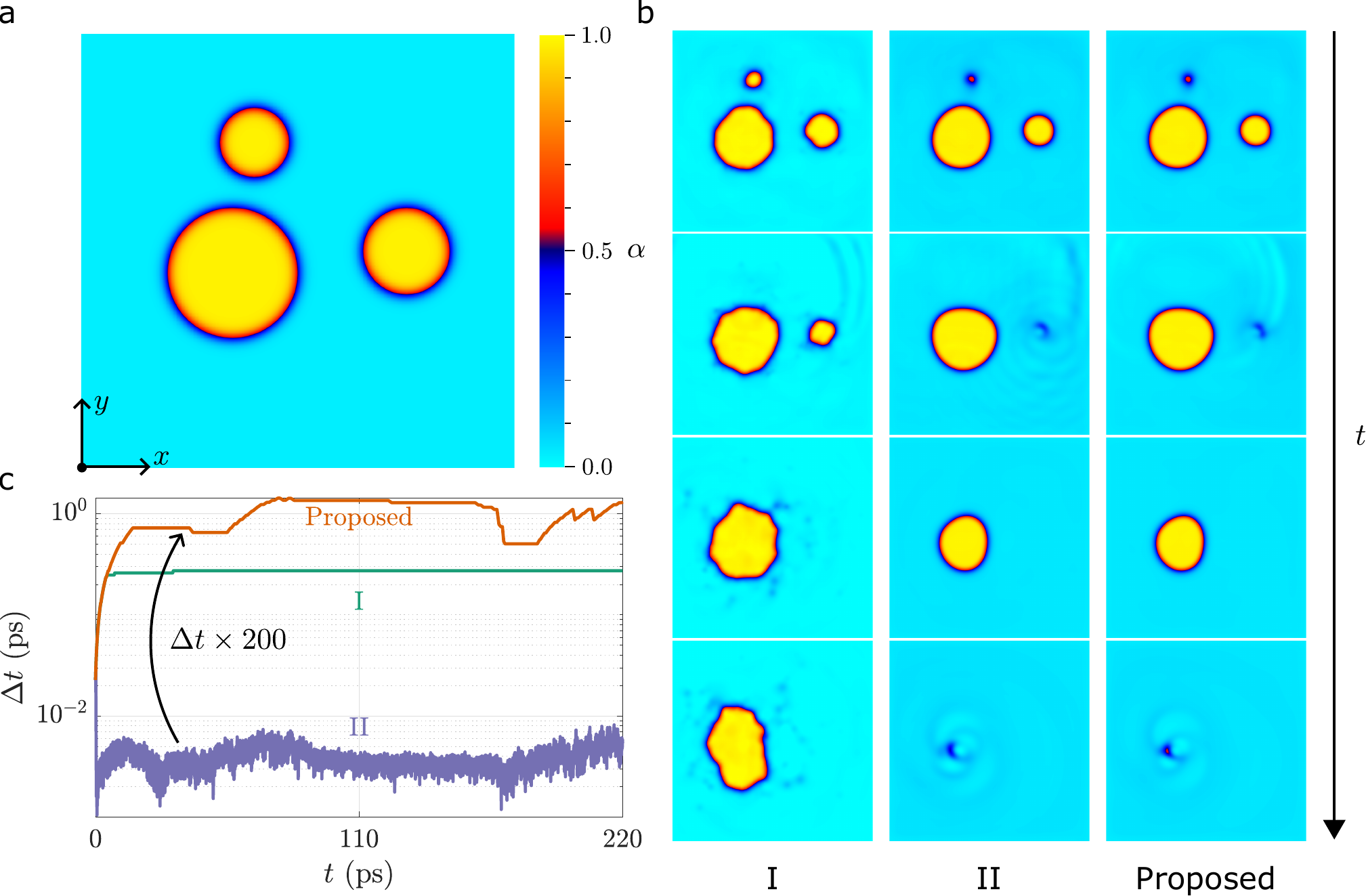}}
  \caption{Stability of the proposed algorithm illustrated by a two-dimensional inviscid simulation of vapor bubble dynamics. The computational domain is a square of side $L_0 = 30$ nm. The mesh is composed of 256$^2$ $C^1-$continuous quadratic elements. An acoustically absorbing layer with a thickness of $0.05L_0$ is placed at the boundary to simulate open boundary conditions. The temperature is $T=550$ K. We use $\lambda = 10^{-16}$ m$^7$/kg/s$^2$ and $\eta = 1$. a. Initial condition representing three vapor bubbles in a liquid pool. The bubble centers are located at $C_1 = (0.25L_0,0.50L_0)$, $C_2=(0.75L_0,0.50L_0)$ and $C_3 = (0.40L_0,0.75L_0)$ while the bubble radii are $R_1 = 0.15L_0$, $R_2 = 0.10L_0$ and $R_3 = 0.08L_0$. b. Instantaneous snapshots of the void fraction $\alpha$ using (I) $\bm{{A}}^{*}_i =  \bm{{A}}^{\rm adv/p}_i + \bm{{A}}^{p}_i$, (II) $\bm{{A}}^{*}_i =  \bm{{A}}^{\rm adv/p}_i + \bm{{A}}^{p}_i - \bm{A}^{c}_i$ and the proposed method; see SUPG operator in \cref{eqn:Astar}. c. Time evolution of the time step for the three different algorithms.}
  \label{fig:SUPG}
\end{figure*}

To compare these three algorithms we simulate the dynamics of three vapor bubbles; see \cref{fig:SUPG}. The expected solution is that the bubbles will collapse one after another from the smallest to the largest. As shown in \cref{fig:SUPG}b (left column), when we use method I, the two smallest bubbles collapse, but the largest bubble acquires an irregular shape and periodically oscillates until the simulation becomes unstable. Algorithm II produces results that, at the scale of the plot, are indistinguishable from those of the proposed algorithm (see central and right columns of \cref{fig:SUPG}b). However, the time step required to get convergence of the Newton-Raphson scheme is $\sim$ 200 times smaller when we use algorithm II than when we employ the proposed method; see \cref{fig:SUPG}c. Overall, this shows that the proposed SUPG operator vastly outperforms naive extensions of the classical SUPG method to the NSK equations.

\begin{figure*}[htbp]%htb
  \centering
  \scalebox{0.8}{\includegraphics{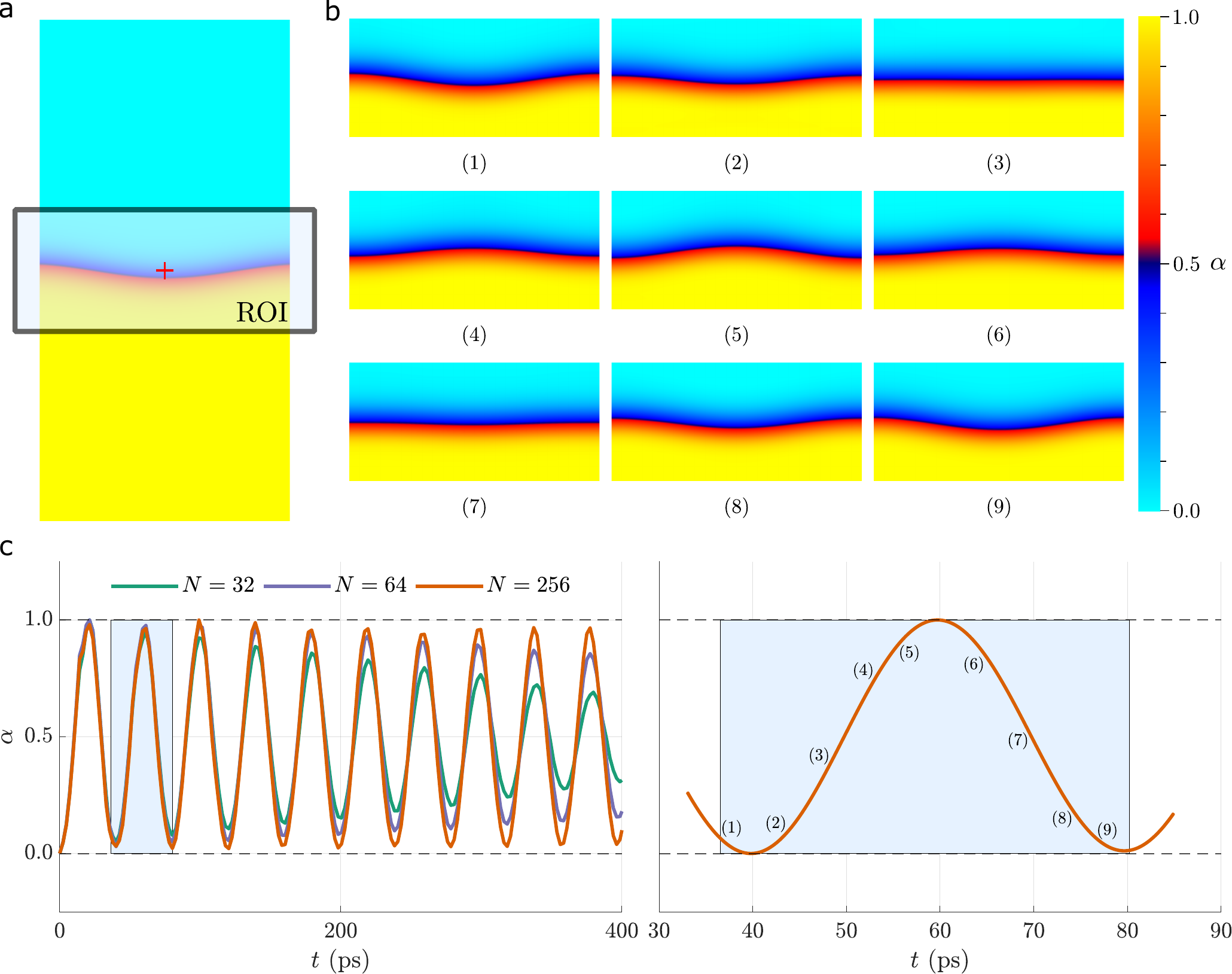}}
  \caption{Accuracy of the proposed algorithm illustrated by a two-dimensional simulation of a liquid-vapor interface oscillation in inviscid flow. The flow temperature is $T=550$ K. We use $\lambda = 10^{-16}$ m$^7$/kg/s$^2$ and $\eta = 1$. We employ free-slip boundary conditions {and $\nabla \rho \cdot \bm{n} = 0$} on the entire boundary. The computational domain is discretized using $N\times 2N$ $C^1-$continuous quadratic elements. {a. Initial condition representing a planar liquid-vapor interface. The initial velocity is zero. b. Snapshots of the void fraction in the region of interest. c. Time evolution of $\alpha$ at the center of the domain for different mesh sizes (left) and inset showing a larger view of the time evolution of $\alpha$ in the second period of oscillation.}}
  \label{fig:Interface}
\end{figure*}

We now perform an additional numerical test to evaluate the numerical dissipation introduced by our algorithm. Most successful algorithms for compressible flows introduce numerical dissipation. However, to obtain an accurate method, the amount of numerical dissipation should quickly approach zero as the mesh is refined. We study the artificial dissipation of our method by simulating the oscillation of an inviscid, planar liquid-vapor interface driven by an initial disturbance. Because the flow is inviscid, we expect a periodic oscillation of the interface without any decay in the amplitude of the disturbance. The initial velocity is zero and the initial void fraction is depicted in \cref{fig:Interface}a. We show snapshots of the void fraction in the region of interest at multiple times in \cref{fig:Interface}b. These pictures show a periodic oscillation of the interface. A more informative description of the oscillation is given by \cref{fig:Interface}c (left), which shows the time evolution of $\alpha$ at the center of the domain for different mesh sizes. The plot shows that the simulation remains stable even for extremely coarse meshes ($N$=32), but there is a significant decay in the wave amplitude due to numerical dissipation. The amplitude decay is less noticeable as we refine the mesh and becomes very small for our finest mesh ($N$=256), even after 10 complete wave periods.

\immediate\write18{texcount -inc -incbib 
-sum Draft03.tex > wordcount1.txt}

\backmatter

%%-----------------------------------------
%% Acknowledgement
%%-----------------------------------------

\bmhead{Acknowledgements}

This work is funded partially by the U.S. Department of Defense (Award No. FA9550-20-1-0165) and partially by National Science Foundation, United States (Award No. 1805817). This work uses the Bridges-2 system %\cite{Brown2021-dr} 
at the Pittsburgh Supercomputing Center (PSC) through allocation \#MCH220014 from the Advanced Cyberinfrastructure Coordination Ecosystem: Services $\&$ Support (ACCESS) program, which is supported by National Science Foundation grants \#2138259, \#2138286, \#2138307, \#2137603, and \#2138296.

%%-----------------------------------------
%% Appendix
%%-----------------------------------------

\begin{appendices}
    \section{Appendix A: Convexity of the Helmholtz free-energy in functional sense}
    \renewcommand{\theequation}{A.\arabic{equation}}

Our model is based on a cubic EoS that is derived from a bulk free energy per unit volume $\psi$ that is non-convex for temperatures lower than the critical temperature; see \cref{eqn:Helmholtz}. However, here we show that, at equilibrium, and under the constraint of mass conservation, the Helmholtz free energy $\mathcal{H}$ is convex in the functional sense even though $\psi$ is not. The constraint of mass conservation can be imposed using a constant Lagrange multiplier \cite{courant2008methods}. For simplicity, we proceed in one spatial dimension and define the constrained Helmholtz free energy as
\begin{equation}
\label{eqn:constrainedH}
    \mathcal{H}_L[{\rho}] = \int _{x_{0}} ^ {x_{1}} \left [\psi(\rho) + \frac{\lambda \eta}{2}(\rho_{,x})^2 - L_m \rho \right ] \dd x,
\end{equation}
where $(x_0,x_1)=\Omega$ is the fluid domain and $L_m$ is the Lagrange multiplier that imposes mass conservation. Extremals of \cref{eqn:constrainedH} represent equilibrium solutions to the NSK equations. Suppose that $\Tilde{\rho}$ is an extremal of $\mathcal{H}_L$. Then, \cref{eqn:constrainedH} can be equivalently written as 
\begin{equation}
\label{eqn:Hlrho}
    \mathcal{H}_L[\Tilde{\rho}] = \int_{x_0}^{x_1} \Psi(\Tilde{\rho},\Tilde{\rho}_{,x},\Tilde{\rho}_{,xx}) \dd x,
\end{equation}
where
\begin{equation}
\label{eqn:Psi}
    \Psi = \psi(\rho) - \frac{\lambda \eta}{2}\rho \rho_{,xx} - L_m \rho.
\end{equation}
To derive \cref{eqn:Hlrho}, we have used the identity
\begin{equation}
    \int_{x_0}^{x_1} \Tilde{\rho} \Tilde{\rho}_{,xx} \dd x = -\int_{x_0}^{x_1} (\Tilde{\rho}_{,x})^2 \dd x + [\Tilde{\rho} \Tilde{\rho}_{,x}]_{x_0}^{x_1},
\end{equation}
which holds because equilibrium solutions correspond to a smooth transition between a uniform vapor phase and a uniform liquid phase that verify $\Tilde{\rho}_{,x}(x_0) = \Tilde{\rho}_{,x}(x_1)=0$; see \cite{Magaletti2015-gi}.

To determine the extremals of $\mathcal{H}_L$, we construct the function $\hat{\rho}(x) = \Tilde{\rho}(x) + \epsilon \varphi(x)$ where $\epsilon$ is a real-valued parameter and $\varphi(x)$ is an arbitrary function that vanishes at $x_0$ and $x_1$. If $\Tilde{\rho}$ is an extremal of $\mathcal{H}_L$, then the function of $\epsilon$ 
\begin{equation}
    \phi(\epsilon) = \mathcal{H}_L[\Tilde{\rho} + \epsilon \varphi]
\end{equation}
must have an extremal at $\epsilon = 0$, that is, $\phi'(0) = 0$. Thus, we define the first variation of $\mathcal{H}_L$ as $\delta \mathcal{H}_L=\phi'(0)$. Using the chain rule we obtain 
\begin{equation}
    \phi'(0) = \int_{x_0}^{x_1} \left [ \Tilde{\Psi}_{,\rho} \varphi + \Tilde{\Psi}_{,\rho_{,x}} \varphi_{,x} + \Tilde{\Psi}_{,\rho_{,xx}} \varphi_{,xx} \right ] \dd x= 0
\end{equation}
where the tildes over the partial derivatives of $\Psi$ indicate evaluation at $\Tilde{\rho}$. To show that $\mathcal{H}_L$ is convex in the functional sense, we need to show that $\mathcal{H}_L$ is minimum at $\Tilde{\rho}$. By using a Taylor expansion of $\phi(\epsilon)$ about $\epsilon = 0$,
\begin{equation}
    \phi(\epsilon)=\phi(0) + \epsilon \phi'(0) + \frac{\epsilon^2}{2} \phi''(\omega); \ \omega \in (0,\epsilon)
\end{equation}
we conclude that $\mathcal{H}_L$ reaches a minimum at $\Tilde{\rho}$ if $\phi(\epsilon)-\phi(0) \ge 0$, which is equivalent to $\phi''(\omega) \ge 0$. Thus, the second variation of $\mathcal{H}_L$ is defined as $\delta^2 \mathcal{H}_L=\phi''(\omega)$. To show that $\delta^2\mathcal{H}_L[\Tilde{\rho}] \ge 0$, we first calculate $\phi''(\omega)$ as
\begin{alignat}{2}
\label{eqn:Phi''}
%\begin{aligned}
    \phi''(\omega) & = \int _{x_0} ^{x_1} \left [ \bar{\Psi}_{,{\rho}{\rho}}\varphi^2 + 2\bar{\Psi}_{,{\rho}{\rho}_{,x}}\varphi \varphi_{,x}  \right ] \dd x \nonumber\\
        & + \int _{x_0} ^{x_1} \left [\bar{\Psi}_{,{\rho}_{,x}{\rho}_{,x}}\varphi_{,x}^2 + 2\bar{\Psi}_{,{\rho}_{,x}{\rho}_{,xx}}\varphi_{,x} \varphi_{,xx} \right ] \dd x \nonumber\\
        & + \int _{x_0} ^{x_1} \left [ \bar{\Psi}_{,{\rho}_{,xx}{\rho}_{,xx}}\varphi_{,xx}^2 + 2\bar{\Psi}_{,{\rho}{\rho}_{,xx}}\varphi \varphi_{,xx}\right ] \dd x,
%\end{aligned}       
\end{alignat}
where the bars over the partial derivatives of $\Psi$ indicate evalution at $\bar{\rho} = \tilde{\rho} + \omega \varphi$. Using \cref{eqn:Psi} and \cref{eqn:Phi''}, it follows that
\begin{equation}
    \delta^2 \mathcal{H}_L[\rho] = \int_{x_0}^{x_1} \varphi \left [ \frac{\partial^2}{\partial \rho^2} \left(\psi(\rho) - L_m \right) -\lambda \eta \varphi_{,xx} \right ] \dd x.
\end{equation}
From the condition $\delta \mathcal{H}_L[\tilde{\rho}] = 0$ we know that
\begin{equation}
\label{eqn:extreme}
    \frac{\partial}{\partial \rho} \left ( \psi(\tilde{\rho})-L_m \right ) - \lambda \eta \tilde{\rho}_{,xx} = 0.
\end{equation}
Exploiting the arbitrariness of $\varphi(x)$, we can take $\varphi(x)=\gamma(x) \tilde{\rho}_{,x}(x)$ which satisfies the only requirement on $\varphi$, i.e., $\varphi(x_0) =\varphi(x_1) =0$. Differentiating \cref{eqn:extreme} with respect to $x$, using $\varphi = \gamma \tilde{\rho}_{,x}$, and performing multiple manipulations, we can show
\begin{alignat}{2}
    %\begin{aligned}
        \delta^2 \mathcal{H}_L[\tilde{\rho}] & =  - \int_{x_0}^{x_1} \lambda \eta  \gamma \Tilde{\rho}_{,x}(2 \gamma_{,x} \tilde{\rho}_{,xx} + \gamma_{,xx} \tilde{\rho}_{,x}) \dd x \nonumber\\
        & = - \int_{x_0}^{x_1} \lambda\eta \gamma \left[ \gamma_{,x} (\rho_{,x})^2 \right]_{,x} \dd x \nonumber\\
        & = \int_{x_0}^{x_1}\lambda \eta (\gamma_{,x})^2 (\tilde{\rho}_{,x})^2 \dd x.
    %\end{aligned}
\end{alignat}
Because $\lambda \eta$ is positive, we conclude that $\delta^2 \mathcal{H}_L[\tilde{\rho}] \ge 0$ and the Helmholtz free energy is convex in the functional sense.

\section{Appendix B: Transformation matrices}\label{sec:Matrices}
\renewcommand{\theequation}{B.\arabic{equation}}
  \begin{equation}
    \bm{A}_{0} =
    \begin{bmatrix} 
      1 & 0 & 0 & 0\\ 
      u_1 & \rho & 0 & 0 \\ 
      u_2 & 0 & \rho & 0 \\ 
      u_3 & 0 & 0 & \rho
    \end{bmatrix},
  \end{equation}

  \begin{equation}
    \bm{A}_{1}^{\rm adv/p} =
    \begin{bmatrix} 
      u_1 & \rho & 0 & 0\\ 
      u_1^2 & 2\rho u_1 & 0 & 0 \\ 
      u_1 u_2 & \rho u_2 & \rho u_1 & 0 \\ 
      u_1 u_3 & \rho u_3 & 0 & \rho u_1
    \end{bmatrix},
    \label{eqn:A_adv_p_1}
  \end{equation}

  \begin{equation}
    \bm{A}_{2}^{\rm adv/p} =
    \begin{bmatrix} 
      u_2 & 0 & \rho & 0\\ 
      u_1 u_2 & \rho u_2 & \rho u_1 & 0 \\ 
      u_2^2 & 0 & 2 \rho u_2 & 0 \\ 
      u_2 u_3 & 0 & \rho u_3 & \rho u_2
    \end{bmatrix},
    \label{eqn:A_adv_p_2}
  \end{equation}

  \begin{equation}
    \bm{A}_{3}^{\rm adv/p} =
    \begin{bmatrix} 
      u_3 & 0 & 0 & \rho\\ 
      u_1 u_3 & \rho u_3 & 0 & \rho u_1 \\ 
      u_2 u_3 & 0 & \rho u_3 & \rho u_2 \\ 
      u_3^2 & 0 & 0 & 2 \rho u_3
    \end{bmatrix},
    \label{eqn:A_adv_p_3}
  \end{equation}

\begin{equation}
    \bm{A}_i^p = p_{,\rho} \bm{e}_{i+1} \otimes \bm{e}_{i}, \text{ (no sum on $i$)}
\end{equation}
\vspace{.1cm}
\begin{equation}
    \bm{A}^c_i = \frac{\lambda \eta \rho \Delta \rho_{,i}}{\rho_{,i}} \bm{e}_{i+1} \otimes \bm{e}_{i}, \text{ (no sum on $i$)}
\end{equation}

  \begin{equation}
    \bm{K}_{11} =
    \begin{bmatrix} 
      0 & 0 & 0 & 0\\ 
      0 & 2\overline{\mu} + \overline{\lambda} & 0 & 0 \\ 
      0 & 0 & \overline{\mu} & 0 \\ 
      0 & 0 & 0 & \overline{\mu}
    \end{bmatrix},
  \end{equation}

  \begin{equation}
    \bm{K}_{12} =
    \begin{bmatrix} 
      0 & 0 & 0 & 0\\ 
      0 & 0 & \overline{\lambda} & 0\\ 
      0 & \overline{\mu} & 0 & 0 \\ 
      0 & 0 & 0 & 0
    \end{bmatrix},
  \end{equation}

  \begin{equation}
    \bm{K}_{13} =
    \begin{bmatrix} 
      0 & 0 & 0 & 0\\ 
      0 & 0 & 0 & \overline{\lambda}\\ 
      0 & 0 & 0 & 0\\ 
      0 & \overline{\mu} & 0 & 0
    \end{bmatrix},
    \label{eqn:K_1i}
  \end{equation}

  \begin{equation}
    \bm{K}_{21} =
    \begin{bmatrix} 
      0 & 0 & 0 & 0\\ 
      0 & 0 & \overline{\mu} & 0 \\ 
      0 & \overline{\lambda} & 0 & 0\\ 
      0 & 0 & 0 & 0
    \end{bmatrix},
  \end{equation}

  \begin{equation}
    \bm{K}_{22} =
    \begin{bmatrix} 
      0 & 0 & 0 & 0\\ 
      0 & \overline{\mu} & 0 & 0\\ 
      0 & 0 & 2\overline{\mu} + \overline{\lambda} & 0 \\ 
      0 & 0 & 0 & \overline{\mu}
    \end{bmatrix},
  \end{equation}

  \begin{equation}
    \bm{K}_{23} =
    \begin{bmatrix} 
      0 & 0 & 0 & 0\\ 
      0 & 0 & 0 & 0 \\ 
      0 & 0 & 0 & \overline{\lambda}\\ 
      0 & 0 & \overline{\mu} & 0
    \end{bmatrix},
    \label{eqn:K_2i}
  \end{equation}

  \begin{equation}
    \bm{K}_{31} =
    \begin{bmatrix} 
      0 & 0 & 0 & 0\\ 
      0 & 0 & 0 & \overline{\mu}\\ 
      0 & 0 & 0 & 0\\ 
      0 & \overline{\lambda} & 0 & 0
    \end{bmatrix},
  \end{equation}

  \begin{equation}
    \bm{K}_{32} =
    \begin{bmatrix} 
      0 & 0 & 0 & 0\\ 
      0 & 0 & 0 & 0\\ 
      0 & 0 & 0 & \overline{\mu}\\ 
      0 & 0 & \overline{\lambda} & 0
    \end{bmatrix},
  \end{equation}

  \begin{equation}
    \bm{K}_{33} =
    \begin{bmatrix} 
      0 & 0 & 0 & 0\\ 
      0 & \overline{\mu} & 0 & 0\\ 
      0 & 0 & \overline{\mu} & 0\\ 
      0 & 0 & 0 & 2\overline{\mu} + \overline{\lambda}
    \end{bmatrix},
    \label{eqn:K_3i}
  \end{equation}
 
\end{appendices}


\bibliographystyle{bst/sn-aps}
\bibliography{mybibfile}% common bib file


\end{document}